\documentstyle[12pt]{article}

\newcommand{\be}{\begin{equation}}
\newcommand{\ee}{\end{equation}}
\newcommand{\ba}{\begin{array}}
\newcommand{\ea}{\end{array}}

\begin{document}
\rightline{LPENSL-TH 09/2001}
\rightline{FNT-T 09/2001}

\begin{center}
{ \Large \bf  Quantum states of elementary three--geometry}

\vspace{24pt}

{\large
{\sl Gaspare Carbone}$\,^{a}$,
{\sl Mauro Carfora}$\,^{b,}$$\,^{1}$
and
{\sl Annalisa Marzuoli}$\,^{b,}$$\,^{2}$}
\vspace{24pt}

$^a$~Ecole Normale Superieure de Lyon,
Laboratoire de Phisique,\\
All\'ee d'Italie 46, 6936 Lyon, Cedex 07 (France)\\
gcarbone@ens-lyon.fr\\

\vspace{12pt}
$^b$~Dipartimento di Fisica Nucleare e Teorica,
Universit\`a degli Studi di Pavia and 
Istituto Nazionale di Fisica Nucleare, Sezione di Pavia, \\
via A. Bassi 6, 27100 Pavia (Italy) \\
$^1$~mauro.carfora@pv.infn.it \\
$^2$~annalisa.marzuoli@pv.infn.it \\

\end{center}

\vspace{12pt}

\begin{center}
{\bf Abstract}
\end{center}

\vspace{12pt}

We introduce a quantum volume operator $K$ in three--dimensional Quantum Gravity by taking into
account a symmetrical coupling scheme of three $SU(2)$ angular momenta.
The spectrum of $K$ is discrete and defines a complete set of eigenvectors
which is alternative with respect to the complete 
sets employed when the usual binary coupling schemes of angular momenta are considered. 
Each of these states, that we call {\em quantum bubbles}, represents an interference
of extended configurations which provides a rigorous meaning to the heuristic notion
of quantum tetrahedron.\\
We study the generalized recoupling coefficients connecting the symmetrical and the binary basis
vectors,  
and provide an explicit recursive solution for such  coefficients by analyzing also its asymptotic limit. 

\vspace{12pt} 

\vfill
\newpage

\section{Introduction}

In addressing three and four--dimensional simplicial quantum gravity models,
we have  to deal with a large number of different formalizations and interpretations which do not seem
yet to converge to a coherent scenario.
For instance, in many approaches the
prominent geometrical objects are not the edge lengths of the triangulations involved -- as in
the original spirit of Regge Calculus \cite{regge} -- but rather area variables, dual graphs or, more generally, 
dual complexes (a critical review on the subject and a list of more recent references can be found in \cite{rewill}).
In such a somewhat confusing situation, a privileged role is still played by the Ponzano-Regge model \cite{ponzano}. 
This is mostly due to the rigorous mathematics that has permeated the subject since the discovery of its regularized counterpart, the Turaev--Viro invariant \cite{turaev}, and to some new interesting geometrical features of 
 the original Ponzano--Regge formula that have been discussed recently in \cite{roberts} and \cite{arc}.\\
As is well known Ponzano--Regge gravity is based on an   
asymptotic formula for the $SU(2)$ $6j$ symbols, {\em i.e.}

\begin{equation}
\{6j\}\;\sim\;\sqrt\frac{1}{12\pi V(T)}\;
\exp \left\{\; i\left( \; \sum_{\alpha=1}^{6}l_{\alpha}\theta_{\alpha} + \pi/4 \right) \right\} \,\doteq \,
\{6j\}_{as}.
\label{asint}
\end{equation}

\noindent Here the limit is taken for all entries $\gg$ 1 and $l=j+1/2$ is the length of an edge in units of $\hbar$
$(j=0,1/2,1,\ldots$). $V(T)$ is the Euclidean volume of the tetrahedron $T$ spanned by the six edges
$\{l_{\alpha}\}$, and finally $\theta_{\alpha}$ is the angle between the outward normals to the faces
which share $l_{\alpha}$.\par
\noindent The physical interpretation of (\ref{asint})
follows if we recognize that the exponential 
includes the classical Regge action (namely the discretized version
of the Hilbert--Einstein action proposed in \cite{regge}) for the tetrahedron $T$. 
On the other hand, the presence of the slowly varying volume term in
front of the phase factor tells us that $\{6j\}_{as}$ can be interpreted as a probability amplitude
in the approach to the classical limit. But, probability amplitude for what? At first sight, for an
elementary block of Euclidean three--geometry to emerge from the recoupling of quantum angular momenta,
or from a {\em spin network} (see also the complementary approach given in \cite{penrose}). 
However, as pointed out in \cite{ponzano} and emphasized in \cite{bieden9}, we can ascribe a physical reality to the
tetrahedron $T$ if and only if its volume --written in terms of the squares of the edges through 
the Cayley determinant-- satisfies the condition $(V(T))^{2}>0$. In other words, the triangle 
inequalities on the four triples of spin variables (associated with the four faces of the tetrahedron)
which ensure the existence of the $6j$ symbol are weaker than the condition that $T$ exists as a
realizable solid.\par
\noindent With these remarks in mind, a few basic questions naturally arise. Why does a classical
three--geometry emerge from a {\em recoupling} of angular momenta? What
does it mean to give the {\em quantum state} of an elementary block of geometry? Which are the
{\em degrees of freedom} of a quantum tetrahedron, if such an object does indeed exist?\par
\noindent In the following we shall show that, by taking 
into account a symmetrical coupling scheme of three $SU(2)$ angular momenta,
new answers to the previous questions arise in a natural way.\par

\section{Quantum bubbles}   

The theory of the coupling of states of three $SU(2)$ angular momenta operators ${\bf J}_1$, ${\bf J}_2$,
${\bf J}_3$ to states of sharp total angular momentum ${\bf J}$ (with projection $J_0$ along
the quantization axis) is 
usually developed in the framework of binary couplings (see {\em e.g.} \cite{bieden9}). Starting from the
ordered triple indicated before, we denote symbolically the admissible schemes according to\par

\begin{equation}
{\bf J}_1+{\bf J}_2={\bf J}_{12}\;\;;\;\;{\bf J}_{12}+{\bf J}_3={\bf J}
\label{j12}
\end{equation}
 
\begin{equation}
{\bf J}_2+{\bf J}_3={\bf J}_{23}\;\;;\;\;{\bf J}_1+{\bf J}_{23}={\bf J}.
\label{j23}
\end{equation}

\noindent The corresponding state vectors may be written as\par

\begin{equation}
|\,[\,(j_1j_2)j_{12}\,j_3]\,jm>\;\doteq\;|j_{12}\,jm>,
\label{st12}
\end{equation}

\begin{equation}
|\,[\,j_1(\,j_2j_3)j_{23}]\,jm>\;\doteq\;|j_{23}\,jm>,
\label{st23}
\end{equation}

\noindent where small letters denote the quantum numbers associated with operators
(more precisely, ${\bf J}_1^2$ $|j_{12}\,jm>$ $=j_1(j_1+1)|j_{12}\,jm>$, {\em etc.} with
the convention $\hbar=1$)
and $m$ is the eigenvalue of $J_0$ with $-j \leq m \leq j$ in integer steps.
Recall that the ket vectors above belong to complete sets defined as 
the simultaneous eigenspaces of the six
commuting operators ${\bf J}_1^2$, ${\bf J}_2^2$, ${\bf J}_3^2$, ${\bf J}^2$, $J_0$ and
${\bf J}_{12}^2$ (${\bf J}_{23}^2$ respectively). Moreover, the elements of such bases
can be expressed as combinations of $\{|j_1m_1>\otimes$ $|j_2m_2>\otimes$ $|j_3m_3>\}$
through suitable (sums over $m$'s of) products of Clebsh--Gordan coefficients.\par
\noindent In this framework the Wigner $6j$ symbol is just the recoupling
coefficient relating the two sets (\ref{st12}) and (\ref{st23}), namely\par

\begin{equation}
<j_{23}\,jm\,|\,j_{12}\,j'm'>\;=
\;\delta_{jj'}\,\delta_{mm'}\,(-1)^{\Phi}\,
 [(2j_{12}+1)\,(2j_{23}+1)]^{1/2}\,
\left\{ \begin{array}{ccc}
j_1 & j_2 & j_{12} \\
j_3 & j & j_{23}
\end{array}\right\},
\label{seij}
\end{equation}

\noindent where $\Phi \equiv j_1+j_2+j_3$. Strictly speaking, the quantum mechanical probability
$|<j_{23}\,jm\,|\,j_{12}\,jm>|^2$ represents the probability that a system prepared in the
state (\ref{st12}) will be found (measured) to be in the state (\ref{st23}). Thus, in spite of
the tetrahedral symmetry of the $6j$ symbol, at the quantum level we cannot recognize
something like a {\em quantum tetrahedron}: it is only in the approach to the classical limit
(\ref{asint}) that both the coupling schemes coexist and the tetrahedron may take shape.\par

This situation changes radically if we bring into play the symmetrical coupling scheme
for the addition of three $SU(2)$ angular momenta to give a fourth definite angular momentum 
${\bf J}$ (with projection $J_0$); symbolically

\begin{equation}
{\bf J}_1+{\bf J}_2+{\bf J}_3\,=\,{\bf J}.
\label{sycoup}
\end{equation}

\noindent As shown independently in \cite{chak} and in \cite{levy} such a democratic coupling is
characterized by the simultaneous diagonalization of the six Hermitian operators
${\bf J}_1^2$, ${\bf J}_2^2$, ${\bf J}_3^2$, $K$, ${\bf J}^2$ and $J_0$, 
where $K$ is a scalar operator built from the irreducible tensor operators
${\bf J}_1$, ${\bf J}_2$, ${\bf J}_3$ and defined according to

\begin{equation}
K\,=\,-\frac{i}{2}\;\epsilon_{pqr}\;
J_{(1)\,p}\,J_{(2)\,q}\,J_{(3)\,r},
\label{volop}
\end{equation}

\noindent where $J_{(\beta)p}$ is the $p$-component of the operator ${\bf J}_{\beta}$ in
spherical coordinates ($p,q,r=$ $+,0,-\;$; $\beta=1,2,3$) and $\epsilon_{+0-}=1$.
Since $K$ is the mixed product of three angular momentum vectors we are entitled to call it {\em volume}.\par
\noindent Each of the eigenvectors of this new set of operators is denoted by\par

\begin{equation}
|\,(j_1j_2j_3)\,k\,jm>\;\doteq\;|k\,jm>,
\label{syst}
\end{equation}

\noindent with  

\begin{equation}
K\,|k\,jm>\;=\;k\,|k\,jm>.
\label{kappa}
\end{equation} 

\noindent Moreover, the set of eigenvalues in each subspace with
$(jm)$ fixed is discrete and consists of pairs $(k,-k)$, with
at most one zero eigenvalue ({\em cfr.} \cite{levy}).\\
From a geometrical point of view a state $|k\,jm>$, for $k\neq 0$,
represents a quantum volume of size $k\;\hbar^3$ which enucleates 
from the symmetrical coupling of three angular momenta 
and $k^2 > 0$ is the quantum counterpart of the condition $(V(T))^2 > 0$ discussed in connection 
with Ponzano--Regge formula (\ref{asint}). As a matter of fact, according 
to the Correspondence Principle in the region of large quantum numbers the states (\ref{syst})
characterize angular momentum vectors confined to narrow ranges around specific values: narrower
are the ranges, the closer we approach the classical regime. This trivially implies that the classical limit
of the operator (\ref{volop}) coincides with the expression of the volume of the classical tetrahedron $T$
spanned by the counterparts of the operators ${\bf J}_1$, ${\bf J}_2$, ${\bf J}_3$. On the other hand,  
since only ${\bf J}^2_1$, ${\bf J}^2_2$, 
${\bf J}^2_3$, ${\bf J^2}$ commute with $K$ at the quantum level, the {\em quantum bubble} $|k\,jm>$ itself
is related to tetrahedral configurations in the sense that the knowledge about 
the two missing edges ${\bf J}_{12}$ and ${\bf J}_{23}$ can be obtained by evaluating the quantum expectation values
of the corresponding operators, {\em i.e.}

\begin{equation}
<k\,jm|\,{\bf J}^2_{12}\,|k\,jm>;\;\;<k\,jm|\,{\bf J}^2_{23}\,|k\,jm>.
\label{expec}
\end{equation}

\noindent Thus we are naturally led to assert that the  Hilbert space spanned by
the quantum bubbles $|k\,jm>$ for fixed $(jm)$ provides a proper 
description of the quantum state of an elementary block of Euclidean three--geometry. 
However, it follows from the above remarks that $|k\,jm>$ cannot be directly interpreted as a 
"quantum tetrahedron", but rather as a quantum interference of extendend configurations
from which information about all significant geometrical quantities can be extracted.\\
At this point it is worthwhile to notice that volume (and area) operators have been 
already introduced in Quantum Gravity.
More precisely, in the framework of the canonical approach based on connections defined 
on an initial data $3$--surface $\Sigma$, such kind of operators have been defined and discussed by many
authors (see \cite{asht}, \cite{rov}, \cite{depi}, \cite{loll}, \cite{barb}). 
In all these approaches the prominent role is played by the kinematical Hilbert
space, namely the space of square--integrable functions defined on the Ashtekar--Isham completion
of $\cal{A}/ \cal{G}$ \cite{ash} (we can think of $\cal{A}/ \cal{G}$ in the present context as the space of 
$SU(2)$--connections on $\Sigma$ modulo gauge tranformations). 
Then a volume operator can be introduced both in the algebraic approach \cite{asht}  
and in the loop quantization scheme  \cite{rov}, \cite{depi}, \cite{loll}, \cite{barb} 
starting from the triad fields, 
which are the variables conjugate to the Yang--Mills connections. 
All such operators are shown to have discrete spectra and can be regularized through suitable prescriptions 
({\em cfr}. also \cite{lewa}, where the relationships among the various approaches are extensively discussed). 
The main drawback is in fact
that in any case the volume functional to be assigned to a region $\subseteq \Sigma$ is not an observable.
Conversely, the volume operator we are dealing with is defined without ambiguites at the quantum level and  
do represent an observable quantity in the basis $|kjm\rangle $. 
Moreover, since $K$ encodes in a natural way a cut-off $\sim\,(k\hbar^{3})^{1/3}$, there is no need for regularization procedures. \\ 
Summarizing the answers to the questions rised in the introduction, we have established that the Hilbert space spanned by the quantum bubbles (\ref{syst}) describes microstates encoding maximized information about their "intrinsic" 
three--dimensional geometry. Furthermore, owing to the presence of the eigenvalue $m$ of $J_0$ as a label of such states, we may assert that there is also a partial information on the orientation of the quantum bubble $|kjm\rangle $ with respect to a fixed quantization axis. Such conclusions are clearly independent of the existence of any underlying theory of gravity, either classical or quantum. Quite intriguing, whenever quantum systems described by states of 
sharp $SU(2)$ angular momenta happen to couple in a symmetrical scheme like (\ref{sycoup}) then a quantum bubble enucleates.\\ 
With these remarks in mind, it is natural to investigate the role that such states may play in (three and four dimensional) quantized theories of gravity based on $SU(2)$ spin networks.\\ 
A preliminary issue concerns the possibility of describing not only a single bubble but rather a 
large portion of quantum space. We may look more closely at the geometry of a quantum bubble by invoking the classical vector diagrams, 
in which the angular momentum operators are depicted as ordinary 3--vectors with suitable indeterminations in their directions. Thus a particular state $|(j_1 j_2 j_3)k\,jm\,\rangle$ 
may be thought of as an oriented quadrilateral --with edges ${\bf J}_1,\, {\bf J}_2,\, {\bf J}_3,\, -{\bf J}$-- which bounds an Euclidean volume $k$. Suppose know we pick up another state  $|(j_1' j_2' j_3')k'\,j'm'\,\rangle$, depending on a different complete set of operators. Obviously the wave functions will be always overlapping, and in principle we could consider any quantum superposition of them. To take advantage of the usual quantization techniques for composite systems we may look at the Hilbert space spanned by the set 

\begin{equation}
|(j_1 j_2 j_3)k\,jm\,\rangle \, \otimes \,  |(j_1' j_2' j_3')k'\,j'm'\,\rangle 
\label{prost}
\end{equation}

\noindent which, for fixed $(jm)$ and $(j'm')$, represents a simultaneous eigenstate of both $K,\, K'$ and 

\begin{equation}
K^{tot}=K\, +\, K'\, .
\label{ktot}
\end{equation}

\noindent If we pretend that the resulting state singles out a maximally connected portion of space --both in the quantum and in the semiclassical regime-- we have to prescribe a rule to join the bubbles. The simplest way to do that is to require that the bubbles glue together along one of their boundary edges. (The situation is quite different from the usual rule adopted in Ponzano--Regge--like approaches, where a three--manifold is built up by assembling  
tetrahedra along pairs of contiguous faces). Going on with our prescription, 
we have to take a suitable contraction of the tensor product in (\ref{prost}) by setting {\em i.e.} 
 
\begin{equation}
{\bf J}_1\, =\, {\bf J}_1' \, .
\end{equation}

\noindent  Whichever way we define such an operation, it is not difficult to recognize that the resulting states are simultaneous eigenvectors of the nine operators 

$$
\left\{ {\bf J}^2_2\, , \, {\bf J}^2_3\, , \,{\bf J}^2\, , \,{\bf J}_0\, , \,{\bf J}'^2_2\, , \,{\bf J}'^2_3\, , \,{\bf J}'^2\, , \,
{\bf J}'_0\, , \, K^{tot} \right\} \, . 
$$

\noindent The resulting configuration --a double bubble--  has six edges in its boundary, 
a partial orientation given by the values $m,\, m'$ and a volume $k^{tot}$ in $\hbar$ units. 
The information about the common edge ${\bf J}_1$ as well as on the partial volumes $k,\, k'$ have 
been wiped out: they can be recovered only by evaluating quantum expectation values. 
If we keep on gluing quantum bubbles, we end up with locally Euclidean three--dimensional spaces (not necessarily  
manifolds) which may (or may not) have a boundary. The loss of information is progressive, and cases in which the final space  is closed and without boundary represent extreme situations, where all geometrical quantities except the eigenvalue of the total volume $K^{tot}$ do not posses  definite values.\\ 
The scenario outlined above could provide a consistent setting for implementing a quantized theory of gravity (or a topological quantum field theory) in (3+1) dimensions by introducing suitable action functionals (or Hamiltonians). However, as will be illustrated in the next section, quite serious technical obstacles in handling even the single quantum bubbles (\ref{syst})  prevent us to formulate more precise
proposals at present.

\section{Generalized recoupling theory}
To get a better insight into the whole subject we turn now to analyze the relation between the different sets of eigenvectors introduced in the previous section. It is worthwhile to note that the $6j$ symbol (\ref{seij}) can be interpreted as a propagator between two states belonging to alternative binary bases. The asymptotic expression (\ref{asint}) is actually the generating functional of a path--sum evaluated at the semiclassical level and the physical information about the underlying classical theory is encoded in the form of the action. \\ 
These remarks suggest us to follow a similar approach in dealing with "symmetrical {\em vs.} binary" coupling schemes. Unfortunately such a generalized recoupling theory has not been studied elsewhere (in \cite{chak} the problem of connecting the symmetrical states to the factorized  Hilbert space $|j_1 m_1 \rangle \otimes |j_2 m_2 \rangle \otimes |j_3 m_3 \rangle $  has been addressed but not completely solved). \\ 
Coming now to the quantitative analysis of the relations between one of the  sets of eigenvectors
arising from a binary coupling ({\em e.g.} the basis (\ref{st12}) parametrized by the eigenvalue of
the intermediate momentum
${\bf J}_{12}$) and the symmetrical basis (\ref{syst}) we introduce 
a unitary transformation defined as\par

\begin{equation}
|k\,jm>\,=\,\sum_{j_{12}}\:\Psi_{j_{12}}^k\:|j_{12}\,jm>.
\label{trans}
\end{equation}

\noindent The generalized recoupling coefficient\par

\begin{equation}
\Psi_{j_{12}}^k\,\doteq\,<(j_1 j_2 j_3)k\,jm\,|\,[(j_1 j_2)\,j_{12}\,j_3]\,jm>\:\equiv
<k\,jm|j_{12}\,jm>
\label{genrec}
\end{equation}

\noindent and its inverse defined as 

\begin{equation}
\Psi^{j_{12}}_k\,\doteq\,<[(j_1 j_2)\,j_{12}\,j_3]\,jm\,|\,
(j_1 j_2 j_3)k\,jm>\:\equiv
<j_{12}\,jm|k\,jm>\,=\,\overline{\Psi_{j_{12}}^k}
\label{invrec}
\end{equation}

\noindent represent the basic ingredients of our discussion. \\ 
We could consider also the second type of binary basis, namely (\ref{st23}), and define the associated
generalized coefficients $\Psi_{j_{23}}^k$ and $\Psi^{j_{23}}_k$. 
However their expressions can be obtained starting 
from (\ref{genrec}) and employing the standard recoupling coefficient given in (\ref{seij}) 
to get \par

\begin{equation}
\Psi_{j_{23}}^k\,=\,\sum_{j_{12}}
\:<k\,jm|j_{12}\,jm>\:
<j_{12}\,jm|j_{23}\,jm>.
\label{rerec}
\end{equation}

\noindent All these coefficients have to be actually interpreted as reduced matrix elements 
since the magnetic quantum numbers can be neglected owing to the Wigner--Eckart theorem. 
Moreover, we do not fix phases and normalizations of such coefficients since there is 
no preferable prescription to do that. \\ 
As pointed out at the end of the previous section the eigenvalue equation (\ref{kappa}) 
for the operator $K$ is quite hard to solve. 
The problem of finding the matrix elements of the operator $K$ in the binary basis
(\ref{st12}) was addressed in \cite{levy}. These authors provided a few  
calculations in the cases $j=j_1+j_2+j_3-\alpha$, with $\alpha=0,1,2$. 
Some progress in generalizing such examples can be actually achieved as explained in 
Appendix A.\\
In Appendix B we show how it is possible to use the
former results to establish a three--term recursion relation 
for the generalized recoupling coefficient (\ref{genrec}), namely

\begin{equation}
k\,\Psi_{\,j_{12}}^k \,=\,i\,\alpha_{\,j_{12}+1}\;\Psi_{\,j_{12}+1}^k \,-\,i\,
\alpha_{\,j_{12}}\;\Psi_{\,j_{12}-1}^k\;.
\label{recur}
\end{equation}

\noindent The functions $\alpha_l$ ($l=j_{12}, j_{12}+1$) can be casted in the form

\begin{equation}
\alpha_l\,=\,[F(l;j_{12}+1/2;j_1+1/2)\,(F(l;j+1/2;j_3+1/2]\,/\,(2l+1)(2l-1),
\label{alfa}
\end{equation}

\noindent where $F(a,b,c)$ represents the area of a triangle with 
edges $a,b,c$: $F(a,b,c)=[(a+b+c)(a-b+c)(a+b-c)(b+c-a)]^{1/2}/4$.\\
Since 

\be
|j_1-j_2|\:\equiv j_{12}^{min}\;\leq\; j_{12} \;\leq\; j_{12}^{ max}\:\equiv j_1+j_2
\label{estremi}
\ee

\noindent we see that the conditions at the extrema fixed by (\ref{recur})
are \par

\begin{eqnarray}
\Psi^k_{j_{12}^{max}} & = & \frac{-i\,\alpha_{j_{12}}^{max}}{k}\:\Psi_{j_{12}^{max}-1}\nonumber\\ 
\Psi^k_{j_{12}^{min}} & = & \frac{i\,\alpha_{j_{12}}^{min}}{k}\:\Psi_{j_{12}^{min}+1}.
\label{extr}
\end{eqnarray}

\noindent It can be shown (see Appendix C) that the formal solution of (\ref{recur}) reads\par
 
\begin{eqnarray}
\lefteqn{\Psi^k_{j^{min}_{12}+n}\;=\;\frac{1}{\prod_{j=1}^{n}(i\alpha_j)}\:\times}\nonumber\\
& & \left[ k^n+\sum_{l=0 }^{[\frac{n}{2}-1]}k^{n-2l-2}
\sum_{\tau_0=2l+1}^{n-1}(i\alpha_{\tau_0})^2
\sum_{\tau_1=2l-1}^{\tau_0-2}(i\alpha_{\tau_{1}})^2\cdots
\sum_{\tau_l=2l-1}^{\tau_{l-1}-2}(i\alpha_{\tau_{l}})^2 \right]\; 
\Psi^k_{j_{12}^{min}}
\label{forsol}
\end{eqnarray}

\noindent where $n$ runs over $\{0,1,2,\cdots,j_1+j_2\}$. As a general remark
we may note that this expression is a real quantity for $n$ even and
purely immaginary when $n$ is odd. The general structure of this 
solution (representing a polynomial of degree $n$ in the variable $k$) is 
quite complicated. On the other hand we make take advantage of having found 
explicitly the generalized recoupling coefficient (\ref{genrec}) in order 
to evaluate expectation values 
of the type (\ref{expec}). More precisely, if $O$ is any 
observable which depends from a set quantum numbers chosen in 
$\{ j_1 , \, j_2 , \,j_3 , \, j , \,  j_{12}\}$ then we could in principle calculate 

$$
\langle k\, jm|\, O \, | k'\, jm \rangle =\sum_{j_{12}} \Psi_{\,j_{12}}^{k'}\, 
\overline{\Psi_{\,j_{12}}^k}\:\:O.
$$

\noindent We can deal also with expectation values of more general observables 
by taking into account combinations of the other recoupling coefficients, namely (\ref{rerec})
(which may be casted into a form similar to (\ref{forsol})) and (\ref{seij}).

\section{Asymptotic limit}

To shed light on the nature of the generalized recoupling coefficient (\ref{genrec})
we can go through a semiclassical (WKB) 
approximation to the recursion relation (\ref{recur}). Such an approach is consistent with 
the Ponzano--Regge analysis (see \cite{ponzano} and \cite{schul}) even if 
in the present case the resulting second order difference equation will turn
out to be much more complicated.\\
The first step consists in modifying the coefficient (\ref{genrec}) into\par 

\begin{equation}
\Phi_{\,j_{12}}^k\,\doteq\,
\Psi_{\,j_{12}}^k\,
[F(h_{12};h_1;h_2)\,F(h_{12};h;h_3)]^{1/2}\,/\,2h_{12},
\label{resc}
\end{equation}

\noindent where $h\equiv j+1/2, h_{12}\equiv j_{12}+1/2$, {\em etc.}.\\
We let now all the $j$-variables to rescale as 

\be
j \:\:\rightarrow\:\: Rj, \:\:R\gg 1 
 \label{jresc}
\ee

\noindent and, by using explicitly the discrete variable $n=j_{12}$ 
$-j_{12}^{min}$ introduced above, we get from (\ref{recur}) a recursion relation
which holds true for the coefficient (\ref{resc}) in the 
semiclassical approximation (\ref{jresc}). It reads \par

\begin{equation}
\Phi_{n+1}^k\,-\,\Phi_{n-1}^k\;=\;i\,C(k,n)\;\Phi_{n}^k,
\label{recas}
\end{equation}

\noindent where \par

\begin{equation}
C(k,\,n)\,=\,2(j_{12}^{min}+n)\,k\,/\,
F(h_{j_{12}^{min}};h_1;h_2)\,F(h_{j_{12}^{min}};h;h_3).
\label{cfunct}
\end{equation}

\noindent If $C(k,\,n)$ were a constant function with respect to the variable $n$
the solution of (\ref{recas}) would go as 

\be
\exp\{\,i\,\arcsin\,[C(k)/2]n\}.
\label{cosol}
\ee

\noindent To deal with the general case we let $n$ become a continuous variable

\be
n\:\rightarrow \:x\:\:\:\:\:\:\mbox{with}\;\;\; x^{min}=0,\: x^{max}=j_1+j_2
\label{contx}
\ee

\noindent and consequently we change the quantities in (\ref{recas}) according to\par

\begin{equation}
\begin{array}{ccc}
\Phi^k_{j_{12}^{min}+n} & \rightarrow & \Phi^k_x \nonumber\\
C(k\,,n) & \rightarrow & C(k\,,x).
\end{array}
\end{equation}

\noindent Thus we can formally associated with (\ref{recas}) an ordinary 
second order difference equation of the type\par

\be
\frac{\Delta^2}{\Delta x^2}\,\Phi^k_x\:+\:2\frac{\Delta}{\Delta x}\,\Phi^k_x\:=\:i\,C(k\,,x)\;\Phi^k_x.
\label{diffeq}
\ee

\noindent
At this point it can be noticed  that the differences between the structure 
of (\ref{diffeq}) and the second order equation which
gives the asymptotic expression (\ref{asint}) for the $6j$ symbol are: 
{\em i)} the $\Phi^k_x$'s are
complex coefficients and formally the substitution $k \rightarrow -k$ changes (\ref{diffeq})
into its complex conjugate; {\em ii)} in our equation a first--derivative term is present too;
{\em iii)} the factor $i\, C(k\,,x)$ is purely immaginary.\\
By taking into account (\ref{cosol}), we search for a solution of (\ref{diffeq}), for each $k$,
in the form\par

\be
\Phi^k_x\:=\:\rho(x)\:\exp\{i\,A(x)\}
\label{rhoa}
\ee

\noindent with $\rho$ and $A$ to be determined. By substitution we get
a pair of differential equations corresponding respectively to the immaginary
and real part of (\ref{diffeq}), namely\par

\be
\begin{array}{c}
2\rho \,\sin A'\,+\,2\rho'\,A''\,\cos A'\,+\,\rho''\,\sin A'\,=\,\rho\, C \nonumber\\
-\rho \,A''\,\sin A'\,+\,2\rho'\, \cos A'\,=\,0.
\end{array}
\label{paireq}
\ee

\noindent Here the primes denote derivatives with respect to the variable $x$ and, 
according to the standard WKB approach, all terms of the type $A''\,\rho''$,
as well as terms containing powers of $A''$, can be assumed negligible 
(these assumptions may be checked {\em a posteriori}).\\
The solutions of (\ref{paireq}) read\par

\be
A(x)\,=\,\int_{x^{min}}^x \,\arcsin \left[ \frac{C(k\,,z)}{2} \right]  dz 
\label{arcsin}
\ee

\be
\rho(x)\,=\,\frac{1}{(\cos A')^{1/2}}\,=\,\left( \sqrt{1-\frac{C(k\,,x)^2}{4}} \right)^{-1/2}. 
\label{rhosolut}
\ee
 
\noindent The first function, which appears under the exponential in (\ref{rhoa}),
can be recasted as\par

\be
A(x)\,=\,\int_{x^{min}}^x \,\arccos \left[ \frac{C(k\,,z)}{2} \right]dz\;+\;\frac{\pi}{2}\,(x-x^{min}) 
\label{arccos}
\ee

\noindent with $(x-x^{min})=n$. In particular, if we compare this expression with the behavior of the 
formal complete solution (\ref{forsol}) with respect to the parity (even/odd powers of $k$),
we see that they agree.\\
Finally, since the classical counterpart of the operator $K$ is the volume $V\equiv V(T)$ of the
Euclidean tetrahedron $T$ we see that the generalized recoupling coefficient (\ref{genrec})
in the asymptotic limit behaves as 

\be
\Psi^k_{j_{12}}\:\sim\:\Psi^V_{j_{12}}\:=\:
\frac{N}{\sqrt{V}} 
\exp \left\{i\,\int_0^{j_{12}} \arccos \left( y\,\sin \theta_{12} \right)
dy\:+\:i\,\pi(j_{12}-j_{12}^{min}) \right\}
\label{gerecas}
\ee

\noindent where $\theta_{12}$ is the dihedral angle between the faces of the tetrahedron $T$
which share $j_{12}$.
We have introduced normalization function $N$ which depends on 
$\{ j_1, \, j_2, \, j_3, \, j, \, k\}$. The determination of such an $N$ is not 
easily performed since we should first normalize the complete coefficient (\ref{forsol})
according to 

$$
\sum_{n}\Psi^k_{j_{12}^{min}+n}\,\overline{\Psi^k_{j_{12}^{min}+n}}\,=\,1 
$$

\noindent For this reasons the lack of symmetry in (\ref{gerecas}) with respect to the other $j$-variables could be eventually recovered. 
Anyway, we may observe that the classical action appearing in the exponential does not look like the Regge action, nor a cosmological--type action.

\section*{Appendix A}

Following \cite{levy} let us consider the matrix elements of the operator $K$ defined in 
(\ref{volop}) with respect to the binary basis (\ref{st12}), namely

\be
<j_{12}\,jm\,|K|\,j'_{12}\,jm>
\label{kj12}
\ee

\noindent If the integer $\nu$ denotes the range of variation of $j_{12}$,
$\nu =2j_{12}+1$, then the $(\nu +1)$ $\times$ $(\nu +1)$ matrix (\ref{kj12})
reads \cite{levy}

\be
{\cal A}_{\nu}\,=\,
\left(\ba{cccccc}
0 & i\alpha_{\nu} & 0 & \cdots & 0 & 0 \\ 
-i\alpha_{\nu} & 0 & i\alpha_{\nu -1} & \cdots & 0 & 0 \\
0 & -i\alpha_{\nu-1} & 0 & \cdots & 0 & 0 \\ 
\vdots & \vdots & \vdots & \cdots & \vdots & \vdots \\
0 & 0 & 0 & \cdots & 0 & i\alpha_1 \\
0 & 0 & 0 & \cdots & -i\alpha_1 & 0 
\ea\right)
\label{matrix}
\ee

\noindent where the explicit expression of the $\alpha$'s is given in (\ref{alfa}).\\
The eigenvalue problem for such a Jacobi--type matrix can be addressed by noticing that 
the characteristic polynomial can be written as \par

\be
{\mbox det}\,({\cal A}_{\nu}-\lambda\, I)\,\doteq
f_{\nu}(\lambda)\,
=\,-\lambda f_{\nu-1}(\lambda)
-(i\alpha_{\nu})^2f_{\nu-2}(\lambda)
\label{carpol}
\ee 

\noindent which has the form of a three--term recursion relation for the functions
$f(\lambda)$.

\vskip 1cm

{\bf Lemma.} The general solution of (\ref{carpol}) is

\begin{eqnarray}
\lefteqn{f_{\nu}(\lambda)\;=\,(-1)^{\nu+1} \times}\nonumber\\
& & \left\{ \lambda^{\nu+1}+ \sum_{k=0}^{[\frac{\nu}{2}-1]}
\lambda^{\nu-2k-1}
\sum_{\tau_0=2k+1}^{\nu}(i\alpha_{\tau_0})^2
\sum_{\tau_1=2k-1}^{\tau_0-2}(i\alpha_{\tau_1})^2
\cdots \sum_{\tau_k=1}^{\tau_{k-1}-2}(i\alpha_{\tau_k})^2 \right\}
\label{eigen}
\end{eqnarray}

{\em Proof.} The proof follows by substitution to each order in $\lambda$.
For $\nu +1$ we get the following expression for the right--hand side of
(\ref{carpol})

\begin{eqnarray}
-\lambda\left\{(-1)^{\nu+1}\left[\lambda^{\nu+1}+\sum_{k=0}^{[\frac{\nu}{2}-1]}
\lambda^{\nu-2k-1}\sum_{\tau_0=2k+1}^{\nu}(i\alpha_{\tau_0})^2
\sum_{\tau_1=2k-1}^{\tau_0-2}(i\alpha_{\tau_1})^2
\cdots \sum_{\tau_k=1}^{\tau_{k-1}-2}(i\alpha_{\tau_k})^2\right]\right\}+\nonumber\\
(i\alpha_{\nu+1})^2\left\{(-1)^{\nu}\left[ \lambda^{\nu}+\sum_{k=0}^{[\frac{\nu}{2}-1]}
\lambda^{\nu-2k-1}\sum_{\tau_0=2k+1}^{\nu}(i\alpha_{\tau_0})^2
\sum_{\tau_1=2k-1}^{\tau_0-2}(i\alpha_{\tau_1})^2
\cdots \sum_{\tau_k=1}^{\tau_{k-1}-2}(i\alpha_{\tau_k})^2\right]\right\}.
\label{solnu}
\end{eqnarray}

\noindent In particular the coefficients of the powers $\lambda^{\nu-2\sigma}$ are

\begin{eqnarray*}
D\,=\,\sum_{\tau_0=2\sigma+1}^{\nu}(i\alpha_{\tau_0})^2
\sum_{\tau_1=2\sigma-1}^{\tau_0-2}(i\alpha_{\tau_1})^2\cdots \nonumber \\
C\,=\,(i\alpha_{\nu+1})^2\sum_{\hat{\tau}_0=2\sigma-1}^{\nu-1}
(i\alpha_{\hat{\tau}_0})^2
\sum_{\hat{\tau}_1=2\sigma-2}^{\hat{\tau}_0-2}(i\alpha_{\hat{\tau}_1})^2\cdots 
\end{eqnarray*}

\noindent and thus the necessary and sufficient condition under which (\ref{solnu})
equals $f_{\nu}(\lambda)$ in (\ref{carpol}) is

\be
D\,+\,C\,\equiv\sum_{\tau_0=2\sigma+1}^{\nu+1}(i\alpha_{\tau_0})^2
\sum_{\tau_1=2\sigma-1}^{\tau_0-2}(i\alpha_{\tau_1})^2\cdots \,=\;E
\label{condns}
\ee

\noindent where $E$ is exactly the coefficient of $\lambda^{\nu-2\sigma}$ in
$f_{\nu}(\lambda)$. 
Then, up to a relabelling $\hat{\tau}_i\rightarrow$ $\tilde{\tau}_{i+1}$ in the function $C$,
(\ref{condns}) is actually fullfilled.$\Box$\\
Analytically we are able to obtain generic solutions only for $\nu \leq 8$, corresponding
to algebraic equations of maximum degree $4$ in $\lambda^2$. These solutions are explicitly listed below

\vskip.5cm

\noindent $\nu=0)\:\:\:\lambda=0$.\\

\noindent $\nu=1)\:\:\:\lambda^2=\alpha_1^2$.\\

\noindent $\nu=2)\:\:\: \lambda^2=\alpha_1^2 +\alpha_2^2; 0$.\\

\noindent By setting $a_{\nu}=\sum_{i=1}^{\nu}\alpha_i^2$ and $b_{\nu}=
\sum_{i=3}^{\nu}\alpha_i^2\sum_{j=1}^{i-2}\alpha_j^2$:\\

\noindent $\nu=3)\:\:\:\lambda^2=\frac{a_3\pm \sqrt{a_3^2-4b_3}}{2}$.\\ 

\noindent $\nu=4)\:\:\;\lambda^2=\frac{a_4\pm \sqrt{a_4^2-4b_4}}{2}; 0$.\\

\noindent By setting $h_{\nu}^2=(\sqrt{(\frac{q_{\nu}}{3})^3+(\frac{p_{\nu}}{2})^2}-\frac{p_{\nu}}{2})^{1/3}
-(\sqrt{(\frac{q_{\nu}}{3})^3+(\frac{p_{\nu}}{2})^2}+\frac{p_{\nu}}{2})^{1/3}$, \\
with $q_{\nu}=b_{\nu}-\frac{a_{\nu}^2}{3}$, $p_{\nu}=\frac{a_{\nu} b_{\nu}}{3}-\frac{2a_{\nu}^3}{27}-c_{\nu}$ and 
$c_{\nu}=\sum_{i=5}^{\nu}\alpha_i^2\sum_{j=1}^{i-2}\alpha_j^2
\sum_{l=1}^{j-2}\alpha_l^2$:\\

\noindent $\nu=5)\:\:\:\lambda_1^2=\frac{a_5}{3}+h_5^2; \, \lambda_2^2=\frac{a_5}{3}
-\frac{h_5^2+\sqrt{h_5^4+4p_5/h_5^2}}{2};\, \lambda_3^2=\frac{a_5}{3}
-\frac{h_5^2-\sqrt{h^4+4p_5/h_5^2}}{2}$.\\ 

\noindent $\nu=6)\:\:\:\lambda_1^2=\frac{a_6}{3}+h_6^2; \, \lambda_2^2=\frac{a_6}{3}
-\frac{h_6^2+\sqrt{h_6^4+4p_6/h_6^2}}{2};\, \lambda_3^2=\frac{a_5}{3}
-\frac{h_6^2-\sqrt{h^4+4p_6/h_6^2}}{2}$.\\

\noindent By setting 
$t_{\nu}^2=2\sqrt{C_{\nu}}-A_{\nu}+2y_{\nu}$, $u_{\nu}^2=y_{\nu}^2+2\sqrt{C_{\nu}}y_{\nu}$,\\ 
$y_{\nu}=(\sqrt{(\frac{Q_{\nu}}{3})^3+(\frac{P_{\nu}}{2})^2}-\frac{P_{\nu}}{2})^{1/3}
-(\sqrt{(\frac{Q_{\nu}}{3})^3+(\frac{P_{\nu}}{2})^2}+\frac{P_{\nu}}{2})^{1/3}$\\
with 
$Q_{\nu}=2C_{\nu}-A_{\nu}\sqrt{C_{\nu}}-\frac{A_{\nu}^2}{12}$, $P_{\nu}=\frac{2C_{\nu}A_{\nu}}{6}
-\frac{A_{\nu}^2\sqrt{C_{\nu}}}{6}-\frac{B_{\nu}^2}{8}-\frac{4A_{\nu}^3}{6^3}$ \\
and 
$A_{\nu}=b_{\nu}-\frac{3a_{\nu}^2}{8}$, $B_{\nu}=\frac{b_{\nu}a_{\nu}}{2}-c_{\nu} -\frac{a_{\nu}^3}{8}$, 
$C_{\nu}=d_{\nu}+\frac{a_{\nu}^2b_{\nu}}{4^2}-\frac{c_{\nu}a_{\nu}}{4}-\frac{3a_{\nu}^4}{4^4}$ \\
with $d_{\nu}=\sum_{i=7}^{\nu}\alpha_i^2\sum_{j=1}^{i-2}\alpha_j^2
\sum_{l=1}^{j-2}\alpha_l^2\sum_{r=1}^{l-2}\alpha_r^2$:

\vskip .5cm

\noindent $\nu=7)\:\:\: \lambda_1^2=\frac{a_7}{4}+
\frac{t_7^2+\sqrt{t_7^4-4(\sqrt{C_7}+y_7-u_7}}{2}); \, 
\lambda_2^2=\frac{a_7}{4}+
\frac{t_7^2-\sqrt{t_7^4-4(\sqrt{C_7}+y_7-u_7}}{2}); \,$\\
$\lambda_3^2=\frac{a_7}{4}-
\frac{t_7^2-\sqrt{t_7^4-4(\sqrt{C_7}+y_7+u_7}}{2}); \,
\lambda_4^2=\frac{a_7}{4}-
\frac{t_7^2+\sqrt{t_7^4-4(\sqrt{C_7}+y_7+u_7}}{2})$.

\vskip .5cm

\noindent $\nu=8)\:\:\:\lambda_1^2=\frac{a_8}{4}+
\frac{t_8^2+\sqrt{t_8^4-4(\sqrt{C_8}+y_8-u_8}}{2}); \, 
\lambda_2^2=\frac{a_8}{4}+
\frac{t_8^2-\sqrt{t_8^4-4(\sqrt{C_8}+y_8-u_8}}{2}); \,$\\
$\lambda_3^2=\frac{a_8}{4}-
\frac{t_8^2-\sqrt{t_8^4-4(\sqrt{C_8}+y_8+u_8}}{2}); \,
\lambda_4^2=\frac{a_8}{4}-
\frac{t_8^2+\sqrt{t_8^4-4(\sqrt{C_8}+y_8+u_8}}{2}); \, 
\lambda_5=0$.

\section*{Appendix B}

Once given the matrix elements (\ref{kj12}) discussed in Appendix A we pass now to consider
the following relation

\be
\sum_{j'_{12}}\;<j_{12}\,jm\,|K|\,j'_{12}\,jm>\;\Psi^k_{j'_{12}}\;=
\;k\, \Psi^k_{j_{12}}
\label{appenb}
\ee

\noindent where $k$ is anyone of the eigenvalues of $K$ according to (\ref{kappa}) 
and $\Psi^k_{j_{12}}$ is the 
generalized recoupling coefficient introduced in (\ref{genrec}).\\
Thus the three-term recursion relation for these coefficients written in (\ref{recur})
is easily found by exploiting the explicit form given in (\ref{matrix}).

\section*{Appendix C}

The recursion relation (\ref{recur}) can be recasted in the form

\be
\Psi^k_{j_{12}^{min}+n}\,=\,
\frac{k}{i\alpha_{j_{12}^{min}+n}}
\:\Psi^k_{j_{12}^{min}+n-1}
\,+\,\frac{i\alpha_{j_{12}^{min}+n-1}}
{i\alpha_{j_{12}^{min}+n}}
\:\Psi^k_{j_{12}^{min}+n-2}
\label{modrec}
\ee

\noindent where $n \in \{0,1,2,\cdots\}$ is defined according to the convention
adopted in the text.\\
This recursion relation and the relation (\ref{carpol}) for the functions
$f(\lambda)$ have the same formal iterative solutions. As a consequence of this remark
we recognize that the expression of the formal solution given in (\ref{forsol})
can be found by following the same steps we used in the Lemma of Appendix A.\\
A few explicit examples of solutions are listed below, where the functions $\alpha$
are given in (\ref{alfa}) and we set

\be
j^{min}_{12}\, \equiv\,\xi.
\ee 

\be
\Psi^k_{\xi+1}\,=\,\frac{k}{i\alpha_{\xi+1}}\Psi^k_{\xi}
\ee

\be
\Psi^k_{\xi+2}\,=\,\frac{1}{(i\alpha_{\xi+1})(i\alpha_{\xi+1})}\:[k^2+(i\alpha_{\xi+1})^2]
\;\Psi^k_{\xi}
\ee

\be
\Psi^k_{\xi +3}\,=\,\frac{1}{(i\alpha_{\xi+1})(i\alpha_{\xi+2})(i\alpha_{\xi+3})}
\{k^3+k[(i\alpha_{\xi+1})^2+(i\alpha_{\xi+2})^2]\}\:
\Psi^k_{\xi}
\ee 

\be
\Psi^k_{\xi+4}\,=\,
\frac{1}{\prod_{j=1}^{4}(i\alpha_{\xi+j})}\:
\left\{k^4+k^2\left[\sum_{j=1}^{3}(i\alpha_{\xi+j})^2\right]+
(i\alpha_{\xi+1})^2(i\alpha_{\xi+3})^2\right\}
\Psi^k_{\xi}
\ee

\begin{eqnarray}
\lefteqn{\Psi^k_{\xi+5}\,=\,\frac{1}{\prod_{j=1}^{5}(i\alpha_{\xi+j})}\times}\hspace{1cm}\nonumber\\
& & \left\{k^5+k^3\left[\sum_{j=1}^{4}(i\alpha_{\xi+j})^2\right]
+k\left[\sum_{j=3}^{4}(i\alpha_{\xi+j})^2\sum_{l=1}^{j-2}(i\alpha_{\xi+l})^2\right]\right\}\,\Psi^k_{\xi}
\end{eqnarray}

\section*{Acknowledgements}

This work is partially supported by MIUR, under PRIN grant 9901457493 ({\em Geometry of
Integrable Systems}).\\ 
We would like to thank V. Aquilanti for very informative and pleasant discussions.

\vfill
\newpage

\section*{References}

\begin{description}

\bibitem[1] {regge} 
Regge T 1961 {\it Nuovo Cimento} {\bf 19} 558

\bibitem[2]{rewill}
Regge T and Williams R M 2000 {\it J. Math. Phys.} {\bf 41} 3964

\bibitem[3] {ponzano}
Ponzano G and Regge T 1968 
{\it Spectroscopic and Group Theoretical 
Methods in Physics} (North-Holland, Amsterdam, F. Bloch et al (Eds.))  p 1

\bibitem[4] {turaev}
Turaev V and Viro O Ya 1992 {\it Topology} {\bf 31} 865

\bibitem[5] {roberts}
J. Roberts 1999 {\it Geometry and Topology} {\bf 3} 21

\bibitem[6] {arc} 
Arcioni G, Carfora M, Marzuoli A and O'Loughlin M 2001  {\it Nucl. Phys.} B {\bf 619} 690

\bibitem[7] {penrose}
Penrose R 1971 
{\it Quantum Theory and Beyond} (Cambridge Univ. Press, T. Bastin (Ed.)) p 151

\bibitem[8] {bieden9}
Biedenharn L C and  Louck J D 1981 
{\it The Racah--Wigner Algebra in Quantum Theory}
(Addison--Wesley, Reading MA) (Topics {\bf 9}, {\bf 12})

\bibitem[9] {chak}
Chakrabarti A 1964 
{\it Ann. Inst. Henri Poincar\'e} {\bf  1}  301

\bibitem[10] {levy}
L\'evy--Leblond J M and L\'evy--Nahas M 1965 {\it 
J. Math. Phys.} {\bf 6} 1372

\bibitem[11] {asht}
Ashtekar A and  Lewandowski J 1997
{\it Adv. Theor. Math. Phys.} {\bf 1} 388

\bibitem[12] {rov} 
Rovelli C and  Smolin L 1995
{\it Nucl. Phys. } {\bf B442} 593 and {\it Nucl. Phys.} B {\bf 456} 743

\bibitem[13] {depi}
De Pietri R and Rovelli C 1996 {\it Phys. Rev} D {\bf 54} 2664

\bibitem[14] {loll}
Loll R 1996
{\it Nucl. Phys.} {\bf  B460} 143

\bibitem[15] {barb} 
Barbieri A 1998 
{\it Nucl. Phys.} B {\bf 518} [PM] 714

\bibitem[16] {ash}
Ashtekar A and Isham C J 1992 {\it Class. Quantum Grav.} {\bf 9} 1433 

\bibitem[17] {lewa} 
Lewandowski J 1997
{\it Class. Quantum Grav.} {\bf 14} 71

\bibitem[18] {schul} 
Schulten K and Gordon R G 1975 
{\it J. Math. Phys.} {\bf 16}  1971

\end{description}

\end{document}